\documentclass[aps,twocolumn,groupedaddress,floatfix,showpacs]{revtex4}
\usepackage{graphicx}
\usepackage{amsmath}

\begin{document}

\newcommand\ket[1]{|#1\rangle}
\newcommand\bra[1]{\langle#1|}
\newcommand\braket[2]{\left\langle#1\left|#2\right.\right\rangle}

\title{Adiabatic pumping in a Superconductor-Normal-Superconductor weak link}

\author{M. Governale$^{1,2}$, F. Taddei$^{1}$, Rosario Fazio$^{1}$, 
and F. W. J. Hekking$^{3}$}
\affiliation{$^1$NEST-CNR-INFM \& Scuola Normale Superiore, I-56126 Pisa, Italy\\
$^2$Institut f\"ur Theoretische Physik III,
Ruhr-Universit\"at Bochum, D-44780 Bochum, Germany\\
$^3$LPMMC, CNRS \& Universit\'e Joseph Fourier, BP 166, 38042 Grenoble CEDEX 9, France} 
\date{\today}
\begin{abstract}
We present a formalism to study adiabatic pumping through a superconductor - normal - 
superconductor weak link. At zero temperature, the pumped charge is related to the 
Berry phase accumulated, in a pumping cycle, by the Andreev bound states.
We analyze in detail the case when the normal region is short compared to the 
superconducting coherence length. The pumped charge turns out to be an even function 
of the superconducting phase difference.
Hence, it can be distinguished from the charge transferred due to the standard
Josephson effect.

\end{abstract}

\maketitle

In a mesoscopic conductor a dc charge current can be obtained, in the absence of
applied voltages, by cycling in time two parameters which characterize the system
\cite{thouless83,switkes99}. This transport mechanism is called pumping. If the time
scale over which the time-dependent parameters vary is large compared to the typical
electron dwell time of the system, then pumping is adiabatic, and the pumped charge
does not depend on the detailed timing of the cycle, but only on its geometrical
properties. Different formulations have been developed to describe adiabatic pumping,
for example, based on scattering theory in Refs.~\cite{brouwer98,makhlin01,buttiker02} 
or Green's function methods in Refs.~\cite{zhou99,entin02}. In the scattering approach 
the pumped charge per cycle can be expressed in terms of
derivatives of the scattering amplitudes with respect to the pumping parameters
(Brouwer's formula \cite{brouwer98}). This result is based on the so-called
emissivities of the system \cite{buttiker94}, which express the charge that flows from
a lead in response to the variation of one parameter in the scattering region. This
formulation requires the presence of terminals which provide propagating channels. The
scattering approach has been later extended to hybrid systems containing
superconducting (S) terminals. In Refs.~\cite{wang01,blaauboer02} a two
terminal structure comprising one superconducting lead was considered. Subsequently,
this approach was generalized to multiple-superconducting-lead systems, where at least
one normal lead is present \cite{taddei04}. The presence of a normal lead is essential
for generalizing Brouwer's formula to these hybrid structures. 

If only superconducting leads are present, at low enough temperature, pumping is due to the adiabatic transport of 
Cooper pairs. Besides the dependence of the pumped charge on the cycle, in the 
superconducting pumps there is a dependence on the superconducting phase difference(s)
(the overall process is coherent). Moreover, in addition to Cooper-pair 
pumping, in equilibrium, there is a contribution due to the Josephson effect 
if the phase difference between the two superconducting leads is different from zero. 
Up to now adiabatic Cooper pair pumping was studied only in the Coulomb Blockade 
regime for all-superconducting systems (superconducting islands weakly connected to superconducting leads)~\cite{geerligs91,pekola99,fazio03,niskanen03}. 
In this Letter we would like to (partially) fill this gap, and consider 
adiabatic pumping between two superconducting terminals connected through a normal (N) 
region.
We focus on the regime of an open 
structure (SNS weak link), where charging effects are negligible.
This is relevant when the normal region is, for example, a chaotic cavity as those used 
for normal, electronic pumps \cite{switkes99}.

The derivation of a formula for the pumped charge makes use of the connection between Berry's 
phase \cite{berry84} and the pumped charge \cite{avron00,aunola03,bender05}, which we prove to be valid also for the SNS weak link.  
The resulting expression for the charge pumped in a period can be written in terms 
of derivatives of the Andreev-bound-state wavefunctions with respect to the pumping parameters.
We point out that there is a close analogy of the problem studied here with that of 
pumping in a  Aharonov-Bohm ring~\cite{moskaletsring}. 

\begin{figure}[t]
    \begin{center}
    \includegraphics[width=\columnwidth,clip]{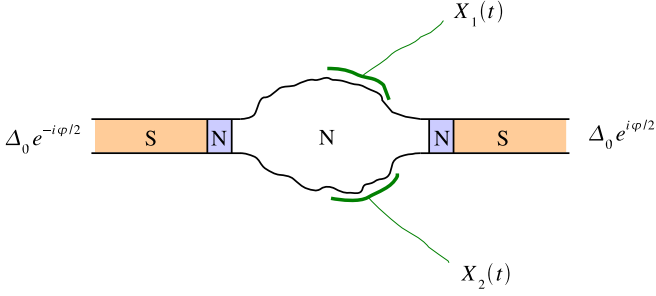}
    \end{center}
    \caption{(Color on line.) Schematic setup for the SNS pump. The weak link is composed 
    	of a scattering region contacted to the S terminals (in orange) through two 
	fictitious N leads (in light blue). The transport properties of the scattering region 
	(characterized by the matrix $S_0$) are cycled in time via two external gates which 
	control the parameters $X_1$ and $X_2$. The superconducting order parameters of 
	the S terminals on the left-hand-side and on the right-hand-side have equal 
	magnitude $\Delta_0$, but differ by a phase $\varphi$.}
    \label{system}
\end{figure}

The system under investigation (depicted in Fig.~\ref{system}) consists of a SNS
junction, with the weak link occupying the region $-W/2<x<W/2$. The superconducting
order parameter is given by $\Delta_0 e^{-i\varphi/2}$ and $\Delta_0 e^{i\varphi/2}$ for
the superconductor on the left-hand-side and right-hand-side, respectively. The
properties of the weak link can be varied, for example, by realizing it with a
semiconductor and operating on two independent external gates, indicated by $X_1(t)$
and $X_2(t)$ in the figure.

The state of such a hybrid structure $\ket{\psi(t)}$ is the solution of the time-dependent 
Bogoliubov-de Gennes equation: 
\begin{equation}
i\hbar \partial_t \ket{\psi(t)}=H(t) \ket{\psi(t)},
\label{BdGeq}
\end{equation}
where the Hamiltonian
\begin{equation}
H(t)=\left(
\begin{array}{cc}
-\frac{\hbar^2}{2m}\vec{\nabla}^2+U(\vec{r},t)-\mu & \Delta (\vec{r})e^{i\phi(\vec{r})} \\
 \Delta (\vec{r})e^{-i\phi(\vec{r})} & \frac{\hbar^2}{2m}\vec{\nabla}^2-U(\vec{r},t)+\mu
\end{array}
\right)
\label{BdGham}
\end{equation}
depends on time through the two parameters: $H(t)=H[X_1(t),X_2(t)]$. In
Eq.~(\ref{BdGham}) $U(\vec{r},t)$ is the potential that takes into account the effect
of the time-varying external gate voltages,
$\phi(\vec{r})=\varphi/2[\theta(x-W/2)-\theta(-x-W/2)]$,
$\Delta(\vec{r})=\Delta_0/2[\theta(x-W/2)+\theta(-x-W/2)]$ and $\mu$ is the
superconductor chemical potential (equal for the two S leads). We now assume that the
state $\ket{\psi(t)}$ evolves adiabatically and that at any time $t$ it is in an
instantaneous eigenstate of the Hamiltonian. The instantaneous solutions are defined
by the equation:
\begin{equation}
H(t)\ket{\tilde{\psi}(t)}=\epsilon(t)\ket{\tilde{\psi}(t)},
\end{equation}
whereby $t$ plays the role of a parameter. After a cycle of period $\tau$, the states returns 
to the initial one,
but with an added phase factor $\Phi$:
\begin{equation}
\ket{\psi(\tau)}=e^{i\Phi}\ket{\psi(0)},
\end{equation}
The phase $\Phi$ contains both  a geometrical (Berry's) and a dynamical
contribution $\Phi=\gamma_{\text{B}}-\gamma_{\text{D}}$. The dynamical
phase is simply given by $ \gamma_{\text{D}}=\frac{1}{\hbar}\int_0^{\tau}dt \, 
\bra{\tilde{\psi}}H\ket{\tilde{\psi}}$

For the SNS system, where Andreev bound states are formed, the condition of validity
of the adiabatic approximation is that the frequency $\hbar \omega$ of the
time-dependent parameters be much smaller
than the energy difference between any pair of Andreev bound states, or between any
Andreev bound states and the continuum of states above the gap. This implies that the
pumping frequency needs at least to be smaller than the superconducting gap
$\Delta_0$.

It is possible to show explicitly that the charge current $J_\text{ch}$ carried by a Bogoliubov-de Gennes eigenstate $\ket{\tilde{\psi}}$ is given by the expectation value of the derivative of the Hamiltonian $H$ with respect 
to the superconducting phase difference \cite{ludin99}:
\begin{equation}
J_\text{ch}=\frac{2e}{\hbar}\bra{\tilde{\psi}}\frac{\partial H}{\partial \varphi}\ket{\tilde{\psi}}.
\label{current}
\end{equation}
The charge transferred per cycle is then defined as $Q=\int_0^{\tau} J_\text{ch}(t) \,
dt.$ By assuming adiabatic evolution of the state, and making use of
Eq.~(\ref{current}), the following relation between the accumulated phase and the
charge transferred in a cycle can be written:
\begin{equation}
Q=2e\frac{\partial}{\partial\varphi}(\gamma_{\text{D}}-\gamma_{\text{B}}).
\label{chfin}
\end{equation}
The first term corresponds to the instantaneous Josephson current integrated over one
period, while the second represents the pumped charge. Using Green's theorem, 
$\gamma_{\text{B}}$ can be written in terms of derivatives with
respect to the pumping parameter of the instantaneous eigenfunctions:
\begin{equation}
\gamma_{\text{B}}=-2\int_{\text{S}} dX_1 dX_2 \text{Im} 
\left[ \braket{\frac{\partial\tilde{\psi}}{\partial X_1}}{\frac{\partial\tilde{\psi}}{\partial X_2}}
\right],
\label{gammaB}
\end{equation}
$S$ being the area in the parameter space spanned by the parameters over one cycle. In
Eq.~(\ref{gammaB}) the notation $\braket{\cdot}{\cdot}$ stands for a space integration
defined by $\braket{A}{B}=\int d\vec{r}\,A^{\dagger}(\vec{r})B(\vec{r})$, $A$ and $B$
being vectors in the Nambu space.

In the short junction limit ({\em i.e.} when the distance $W$ between the two
superconducting interfaces is much smaller than the superconducting coherence length)
only the superconducting regions contribute to the space integration in
Eq.~(\ref{gammaB}).
The instantaneous eigenfunction, corresponding to the Andreev-bound-state energy $\epsilon_j$, in the superconducting regions
can be written as:
\begin{equation}
\tilde{\psi_j}(\vec{r})=\left\{ \begin{array}{lr} \tilde{\psi}_{\text{S,L},j}(\vec{r}) & 
x\le -W/2 \\\tilde{\psi}_{\text{S,R},j}(\vec{r}) & x\ge W/2 \end{array}
\right. ,
\end{equation}
with
\begin{equation}
\tilde{\psi}_{\text{S,L},j}(\vec{r})=\sum_n\left( \mathbf{b}^{+}_{\text{L}n,j}
e^{-ik^+_{n,j} x} +\mathbf{b}^{-}_{\text{L}n,j} e^{ik^-_{n,j}x} \right)\chi_{n}(y,z)
\label{psiS1}
\end{equation}
and
\begin{equation}
\tilde{\psi}_{\text{S,R},j}(\vec{r})=\sum_n\left( \mathbf{b}^{+}_{\text{R}n,j} 
e^{ik^+_{n,j}x} +\mathbf{b}^{-}_{\text{R}n,j} e^{-ik^-{_n,j}x} \right)\chi_{n}(y,z),
\label{psiS2}
\end{equation}
$n$ being the transverse channel index relative to the transverse wavefunction
$\chi_n(y,z)$ with transverse subband energies $E_n$ \cite{note0}. 
The index $j$ labels the different Andreev bound states.
In Eqs.~(\ref{psiS1}) and (\ref{psiS2}) $k^{\pm}_{n,j}$ are particle(hole)-like quasiparticle 
wavevectors given, in the Andreev approximation 
($\Delta_0\ll\mu-E_n$ for any $E_n$), 
by $k^{\pm}_{n,j}= 
k_{\text{F}n} \left(1\pm
\frac{i}{2}\sqrt{\frac{\Delta^{2}_{0}-\epsilon_j^{2}}{(\mu-E_n)^2}}\right)$, with
$k_{\text{F}n}=\sqrt{\frac{2m}{\hbar^2} (\mu-E_n)}$.
The Nambu-space vectors
$\mathbf{b}^{\beta}_{\nu n,j}$ can be calculated with the following procedure:
i) the
eigenfunction in the fictitious leads in the normal regions adjacent to the
superconducting interface (see Fig. \ref{system}) are calculated along the lines of
Ref.~\cite{beenakker92}; ii) the wavefunction in the superconductor is obtained by
imposing the continuity equations at the interfaces within the Andreev approximation;
iii) the normalization condition $\sum_{\nu=\text{L,R}}
\braket{\tilde{\psi}_{\text{S},\nu,j}}{\tilde{\psi}_{\text{S},\nu,j}}=1$ is imposed.

Making use of the Andreev approximation, the pumped
charge reduces to
\begin{widetext}
\begin{equation}
\label{central}
Q_{\text{p
}}=4 e \frac{\partial}{\partial \phi}\int_{\text{S}}dX_1 dX_2 \sum_{\stackrel{{\stackrel{\beta=+,-}
{\nu=\text{L,R}}}}{n,j}}\left\{
\frac{1}{2\text{Im}k^{+}_{n,j}}
\text{Im}\left[\frac{\partial \mathbf{b}_{\nu n,j}^{\beta\dagger}}{\partial X_1}
\frac{\partial \mathbf{b}_{\nu n,j}^{\beta}}{\partial X_2}\right]+
\frac{1}{(2\text{Im}k^{+}_{n,j})^2}
\text{Re}\left[ \mathbf{b}^{\beta\dagger}_{\nu n,j}
\frac{\partial \mathbf{b}^{\beta}_{\nu n,j}}{\partial X_2}  \frac{\partial k^{+}_{n,j}}{\partial X_1}+
\frac{\partial \mathbf{b}^{\beta\dagger}_{\nu n,j}}{\partial X_1} 
\mathbf{b}^{\beta}_{\nu n,j}\frac{\partial k^{+}_{n,j}}
{\partial{X_2}}
\right]\right\},
\end{equation}
\end{widetext}
where the sum over $j$ runs over the Andreev bound states.  
This is the central result of this Letter, and a few comments are in order. We have
succeeded in expressing the pumped charge as a function of the instantaneous
Andreev-bound-state eigenfunction. The vectors
$\mathbf{b}^{\pm}_{\text{L(R)}n,j}$ depend only on the parameters of the system,
such as the normal region scattering matrix $S_0$, the superconducting gap, and the
superconducting phase difference. It is clear that the pumped current can be written
in terms of the elements of the normal-region scattering matrix $S_0$.
Equation~(\ref{central}) is a zero temperature result, and contains only the
contribution to the pumped charge due to the Andreev bound states. 
At finite temperatures, but still smaller than the gap, the contributions of the different Andreev bound 
states $\epsilon_j$ are weighted by the thermal occupation $1-2 f(\epsilon_j)$, being $f$ the Fermi function. 
At temperatures of the order of the gap, there is an additional contribution, 
not contained in Eq. (\ref{central}), due to the propagating quasi particles with energies 
above the gap. When superconductivity is suppressed only the latter contribution, 
which is described by Brouwer's formula, remains, leading to the usual result for 
the pumped charge through a normal region connected to normal leads.

Now let us consider the following single-channel parametrization for the
normal-conductor scattering matrix
\begin{equation}
S_{0}=\left(
      \begin{array}{cc} e^{i \alpha} \sqrt{1-g} & i \sqrt{g}\\
          i \sqrt{g} &  e^{-i \alpha} \sqrt{1-g}
\end{array}\right),
\end{equation}
choosing $g$ and $\alpha$ as pumping fields ($X_1=g$, and $X_2=\alpha$).
The instantaneous Andreev-bound-state energy $\epsilon_0$ is simply related to the 
transmission probability $g$ by \cite{beenakker92}
\begin{equation}
\epsilon_0(t)=\Delta_0\sqrt{1-g(t)\sin^2(\frac{\varphi}{2})},
\end{equation}
so that $\gamma_{\text{D}}=1/\hbar\int_0^{\tau} \epsilon_0(t)$.

The charge transfered due to the Josephson current (in the following named also 
Josephson charge) reads
\begin{equation}
Q_{\text{Jos}}=- 2 e \frac{\Delta_0}{\hbar}
\int_0^{\frac{2\pi}{\omega}} dt\,\,\,
\frac{g(t)}{2}\frac{\sin{\left(\frac{\varphi}{2}\right)}\cos{\left(\frac{\varphi}{2}\right)}}
{\sqrt{1-g(t)\sin^2{\left(\frac{\varphi}{2}\right)}}}.
\label{jJos}
\end{equation}
Notice that it depends on the pumping frequency $\omega$.
On the other hand,
the  pumped charge does not depend on the pumping frequency,
but only on the geometrical properties of the pumping cycle, and it
reads
\begin{equation}
Q_{\text{p}}=4 e\int_S dg d\alpha \frac{\partial}{\partial\varphi} \text{Im} 
\left[ \braket{\frac{\partial\tilde{\psi}}{\partial g}}{\frac{\partial\tilde{\psi}}
{\partial \alpha}} \right].
\label{jpump}
\end{equation}
Interestingly, the integrand of Eq.~(\ref{jpump}) turns out to be independent of
$\alpha$.

We now consider the following sinusoidal pumping cycle: $g(t)=\bar{g} +\Delta g
\sin(\omega t)$ and $\alpha(t)=\bar{\alpha} +\Delta\alpha \sin(\omega t+\phi_0)$. In
the weak pumping limit we assume that $\Delta g/\bar{g}\ll 1$ so that the integrand of
Eqs.~(\ref{jJos}) and (\ref{jpump}) vary negligibly during the cycle. As far as the
frequency is concerned, the maximum value of $\hbar\omega$ in order for the adiabatic
hypothesis to hold is $\hbar\omega<\Delta_0-\bar{\epsilon}_0$, with
$\bar{\epsilon}_0=\Delta_0\sqrt{1-\bar{g}\sin^2(\varphi/2)}$. In order to compare
$Q_{\text{p}}$ with $Q_{\text{Jos}}$, we choose
$\hbar\omega=0.1(\Delta_0-\tilde{\epsilon}_0)$, $\tilde{\epsilon}_0$ being equal to
the value of $\bar{\epsilon}_0$ at $\varphi=\pi/2$. Note that the adiabatic condition
breaks down for $\varphi=0$ or $\bar{g}=0$, when the Andreev bound state is at the gap
boundary. Thus in our analysis we shall avoid small values of those variables.

\begin{figure}
    \begin{center}
    \includegraphics[width=\columnwidth,clip]{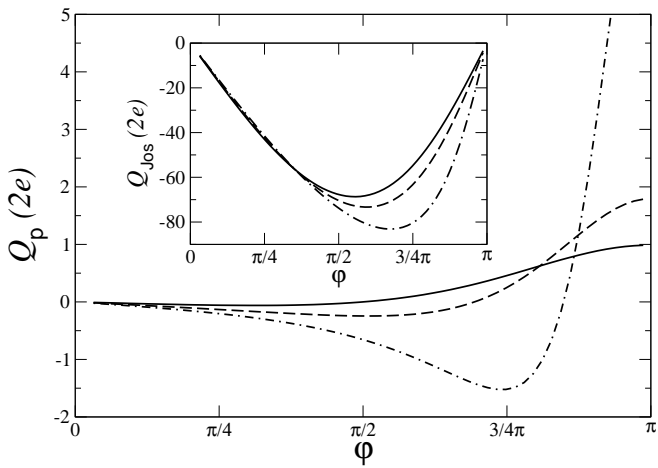}
    \end{center}
    \caption{Pumped charge, in units of $2e$, as a function of superconducting phase 
difference $\varphi$ computed for a sinusoidal weak-pumping cycle. Different lines 
refer to different values of average transmission: solid line $\bar{g}=0.5$; dashed 
line $\bar{g}=0.7$; and dot-dashed line $\bar{g}=0.9$. For the pumping cycle we have 
chosen $\Delta g=0.1$, $\Delta\alpha=2\pi$ and $\hbar\omega=0.1\Delta_0(1-\sqrt{1-\bar{g}/2})$. 
In the inset the Josephson charge is plotted as a function of $\varphi$ for the same 
parameters of the main panel.}
    \label{Qvsfi}
\end{figure}

Figure~\ref{Qvsfi} shows the pumped charge as a function of the
superconducting phase difference $\varphi$ for different values of $\bar{g}$.
$Q_{\text{p}}$ is a non-monotonous function of $\varphi$ exhibiting a maximum at $\varphi=\pi$.
For comparison, in the inset of Fig. \ref{Qvsfi}, we plot the transferred charge due 
to the Josephson current, whose absolute value is larger, with respect to the pumped 
current, by a factor of order $\hbar\omega/\Delta_0$.
However, while $Q_{\text{p}}(\varphi)$ is an even function of $\varphi$, 
$Q_{\text{Jos}}(\varphi)$ is odd, so that a measure of $[Q(\varphi)+Q(-\varphi)]/2$ 
will single out only the pumped contribution.
The particular parity of $Q_{\text{p}}$ is due to the fact that a time-reversal 
operation implies not only the reversal of phase but also of the pumping 
trajectory in parameter space.
It has also to be noticed (see Fig. \ref{Qvsfi}) that the pumped charge is not quantized.
In absence of Coulomb blockade, charge quantization occurs only for
very specific pumping cycles (see, for example, Ref.
\cite{makhlin01}). In addition the global superconducting
phase-coherence of the system, which leads to wave functions
extending in the two superconducting electrodes, also hinders charge
quantization.

To study the effect of an external magnetic field in the normal region,
we change slightly the parametrization of $S_0$ introducing a phase factor $\exp{\pm i\beta}$ 
in the transmission amplitudes.
As a result both the pumped and the Josephson charge are shifted along
the $\varphi$ axis by $2\beta$, i.e. $Q_{\text{p/Jos}}(\varphi)\rightarrow J_{\text{p/Jos}}
(\varphi-2\beta)$. For example, the maximum pumped charge is now reached for $\varphi=2\beta-\pi$.

As far as detection is concerned, we wish to stress that for realistic
parameters, using Al as superconductor,
one can attain sizable pumped currents of the order of 5 nA.
A sensitive setup to currents of even smaller size can be realized by inserting the SNS 
pump in a arm of a SQUID.

In conclusion, we have presented a formalism to study adiabatic charge pumping
in a SNS weak link. The pumped charge is related to the Berry's phase accumulated in a 
pumping cycle by the Andreev bound-state wavefunctions, which can be written as a function 
of the scattering matrix of the normal region. 
In the short junction limit, the pumped
charge is even with respect to the superconducting phase difference. Hence, it can be 
easily distinguished from the charge transferred by the Josephson current.

\begin{acknowledgments}
We acknowledge support from Institut Universitaire de France (F.W.J.H.)
and from EC through grants EC-RTN Nano, EC-RTN Spintronics and EC-IST-SQUIBIT2
(M.G., F.T. and R.F.).
\end{acknowledgments}


\begin{thebibliography}{24}
\bibitem{note0} Equations (\ref{psiS1}) and (\ref{psiS2}) are valid in general. However, beyond the short-junction limit the knowledge of the wave function in the normal region is necessary to compute the space integration in Eq.~(\ref{gammaB}).
\end{thebibliography}
\end{document}